\documentclass[manuscript,screen]{acmart}

\AtBeginDocument{%
  \providecommand\BibTeX{{%
    \normalfont B\kern-0.5em{\scshape i\kern-0.25em b}\kern-0.8em\TeX}}}

\setcopyright{acmcopyright}
\copyrightyear{TBC}
\acmYear{TBC}
\acmDOI{TBC}

\acmPrice{15.00}
\acmISBN{978-1-4503-XXXX-X/18/06}

\begin{document}

\title{rTsfNet: a DNN model with Multi-head 3D Rotation and Time Series Feature Extraction for IMU-based Human Activity Recognition }
\renewcommand{\shorttitle}{rTsfNet}

\author{Yu Enokibori}
\email{enokibori@i.nagoya-u.ac.jp}
\orcid{0000-0001-7350-8866} 
\affiliation{%
  \institution{Graduate School of Informatics, Nagoya University}
  \streetaddress{\#361, IB-south-tower, Nagoya University, Furo-cho, Chikusa-ku}
  \city{Nagoya}
  \state{Aichi}
  \country{Japan}
  \postcode{464-8603}
}

\renewcommand{\shortauthors}{Yu Enokibori}

\begin{abstract}

Although many deep learning (DL) algorithms have been proposed for the IMU-based HAR domain, traditional machine learning that utilizes handcrafted time series features (TSFs) still often performs well. It is not rare that combinations among DL and TSFs show better accuracy than DL-only approaches. However, there is a problem with time series features in IMU-based HAR. The amount of derived features can vary greatly depending on the method used to select the 3D basis. Fortunately, DL's strengths include capturing the features of input data and adaptively deriving parameters. Thus, as a new DNN model for IMU-based human activity recognition (HAR), this paper proposes rTsfNet, a DNN model with Multi-head 3D Rotation and Time Series Feature Extraction. rTsfNet automatically selects 3D bases from which features should be derived by extracting 3D rotation parameters within the DNN. Then, time series features (TSFs), based on many researchers' wisdom, are derived to achieve HAR using MLP. Although rTsfNet is a model that does not use CNN, it achieved higher accuracy than existing models under well-managed benchmark conditions and multiple datasets: UCI HAR, PAMAP2, Daphnet, and OPPORTUNITY, all of which target different activities.

\end{abstract}




\maketitle

\section{Introduction}
Since the advent of deep learning (DL), many DL methods have been proposed for IMU-based human activity recognition (HAR). Many related mechanisms also have been proposed frequently, such as HAR dataset federation using DNN \cite{10.1145/3550285}, self-supervised learning for HAR \cite{10.1145/3550299, 10.1145/3550319}, and so on. Focusing on their core recognition part, many of them are influenced by the image processing field and apply CNN to IMU output values or RNN to CNN values to derive features and discriminative results without handcrafted time series features. 

However, in the IMU-based HAR domain, traditional machine learning (ML) that utilizes handcrafted time series features, which are comprised of the intellectual wisdom of many researchers, still often performs well. For example, as shown in Table \ref{tab:UCIHAR-comp}, the reference result of the UCI HAR \cite{ucihar} Benchmark setup that uses SVM and 561 handcrafted features has an accuracy of 96.37, a value that exceeds almost DL-based approaches. The second-ranked ML is that of C.A. Ronao et al. \cite{doi:10.1177/1550147716683687} whose approach uses Hierarchical Continuous HMMs. Their approach shows an accuracy of 93.18 under the UCI HAR Benchmark setup, which is only 2 points lower than the DL of 95.40 with CNN and Attention by Z.N. Khan et al. \cite{KHAN2021107671} In addition, methods that combine CNN and handcraft features, such as FFT, as proposed by A. Ignatiov \cite{IGNATOV2018915}, W. Jiang \cite{10.1145/2733373.2806333}, and so on., achieved higher accuracy than methods using the Attention mechanism, which is one of the newest DL approaches. The above facts indicate that handcraft-based time series features (TSF) may suit IMU-based HAR.

\begin{table}[tb]
\centering
\caption{Comparition on UCI HAR}
\label{tab:UCIHAR-comp}
\begin{tabular}{l l l l l l l}
\toprule
study & year & note & acc & mf1 & wf1 & note\\
\midrule
\anon[Author(s)]{Y. Enokibori} & - & proposed & 97.76 & 0.9779 & 0.9776 & - \\
A. Ignatov \cite{IGNATOV2018915} & 2018 & CNN + stat. f. + data-centering  & 97.63 & - & - & use test \\
& & & & & & set info. \\
W. Jiang et al.\cite{10.1145/2733373.2806333} & 2015 & 2D FFT + CNN + SVM & 97.59 & - & - & ensemble \\
\anon[Author(s)]{Y. Enokibori} & - & proposed, w/o mh-3D-rotation & 96.71 & 0.9675 & 0.9670 & - \\
L. Lu et al. \cite{9802089}& 2022 & CNN + GRU & 96.67 & - & - & - \\
D. Anguita et al.\cite{ucihar} & 2013 & SVM + 561 hc. features & 96.37 & 0.9638 & 0.9636 & reference \\
A. Ignatov \cite{IGNATOV2018915} & 2018 & CNN + stat. feature & 96.06 & - & - & - \\
C.A. Ronao et al.\cite{RONAO2016235} & 2016 & FFT + CNN & 95.75 & - & - & - \\
Z.N. Khan et al.\cite{KHAN2021107671} & 2021 & Attention + CNN & 95.40 & - & - & - \\
A. Dahou et al.\cite{DAHOU2022111445}& 2022 & DNN + BAOA + SVM & 95.33 & 0.9533 & - & ensemble \\
Y. Dong et al.\cite{dong:hal-02873462} & 2020 & DSMT & 95.31 & 0.9467 & - & - \\
W. Jiang et al.\cite{10.1145/2733373.2806333} & 2015 & 2D FFT + CNN & 95.18 & - & - & - \\
K. Xia et al.\cite{9043535}& 2020 & LSTM-CNN & 95.14 & 0.9525 & - & - \\
J. He et al.\cite{8570822} & 2019 & SVL + RNN + Attention & 94.80 & - & - & - \\
C.A. Ronao et al.\cite{RONAO2016235} & 2016 & CNN & 94.79 & - & - & - \\
C. Xu et al.\cite{8598871} & 2019 & CNN + LSTM & 94.60 & - & - & - \\
N. Tufre et al.\cite{8918509} & 2020 & LSTM & 93.70 & 0.8260 & - & - \\
Y. Zhao et al.\cite{zhao2018deep} & 2018 & Bi-LSTM & 93.60 & - & - & - \\
K. Wang et al.\cite{8716726} & 2019 & Attention + CNN & 93.41 & - & - & - \\
C.A Ronao\cite{doi:10.1177/1550147716683687}& 2017 & Hierarchical Continuous HMM & 93.18 & - & - & - \\
V. Bianchi et al.\cite{8727452} & 2019 & CNN & 92.50 & 0.9210 &  & - \\
Y. Li et al.\cite{10.1007/978-3-319-13817-6_11} & 2014 & Stacked Auto Encoders + SVM & 92.16 & - & - & - \\
Y. Li et al.\cite{10.1007/978-3-319-13817-6_11} & 2014 & PCA + SVM & 91.82 & - & - & - \\
C.A. Ronao et al.\cite{6975918} & 2014 & HMM & 91.76 & - & - & - \\
H. Li et al.\cite{8861371} & 2020 & Bi-LSTM & 91.21 & 0.9001 & - & - \\
C.A. Ronao et al.\cite{10.1007/978-3-319-26561-2_6} & 2015 & CNN & 90.89 & - & - & - \\
S. Seto et al.\cite{7376775} & 2015 & DTW & 89.00 & - & - & - \\
D. Anguita et al.\cite{10.1007/978-3-642-35395-6_30} & 2012 & Handcrafted features + SVM & 89.00 & - & - & - \\
Y.J. KIM et al.\cite{7379660} & 2015 & HMM & 83.51 & - & - & - \\
\bottomrule
\end{tabular}
\end{table}

On the other hand, there is a problem with time series features in IMU-based HAR. The amount of derived features can vary greatly depending on the method used to select the 3D basis. For example, even if a sensor is mounted at the same wrist position, different features will be derived if the mounting angle is different. This problem is commonly solved by extracting an effective 3D basis using PCA. The use of the L2 norm is also often used to ignore those basis differences. However, selecting only one 3D basis by PCA is undeniably inadequate for HAR, which is conducted under various conditions. In addition, this approach is not good at selecting axes by comprehensively considering multiple sensors, such as accelerometers, gyroscopes, and magnetometers. Only using the L2 norm tends to reduce the amount of information, which in turn lowers the final accuracy limit.

Fortunately, DL's strengths include capturing the features of input data and adaptively deriving parameters. Thus, this paper proposes rTsfNet, which selects multiple 3D bases in a DNN and derives time series features from the data in them. Although rTsfNet is a model that does not use CNNs, its accuracy still outperforms existing models under multiple benchmark setups.

The remainder of this paper is organized as follows: Section \ref{sec:rw} describes the verifiability issue of IMU-based HAR and explores related studies based on well-managed benchmark setups. Our proposed model, rTsfNet, is discussed in Section \ref{sec:rtsfnet}. Section \ref{sec:eval} evaluates and compares its performance with other studies. After rTsfNet's potential is described in Section \ref{sec:potential}, this paper's conclusion is presented in Section \ref{sec:conclusion}.

\section{Discussion about IMU-based HAR Benchmark and studies}\label{sec:rw}
This section summarizes the current status of IMU-based HAR. First, we discuss the verifiability issue as a prerequisite for comparing IMU-based HAR studies. Then, we summarize related studies based on the UCI HAR Benchmark setup, which has a clean evaluation setup, allows direct comparison, and has been used in many studies. Note that this paper does not discuss ensemble learning, data augmentation, semi-supervised learning, and other techniques that extend basic algorithms.

\subsection{Verifiability issue}
\subsubsection{Direct comparability issue with benchmark setup}
Many datasets for IMU-based human action recognition have been proposed \cite{ucihar, pamap2, daphnet, opportunity, wisdm, realworld}. Many IMU-based HAR studies have evaluated their methods and compared them with other methods on identical datasets. However, more than just using the same dataset is required to properly compare action recognition methods properly. The segmentation and train/test set split methods must be unified. 

For the segmentation method, the window and sliding sizes must be clear, at least for the test set. For example, the difficulty of detecting a gait activity with 0.5 seconds of window size data differs from detecting such activity with 10 seconds. 

The train/test set split must be subject-based, trial-based, or time-based. Simple random ratio-based splitting is not acceptable in IMU-based HAR because the data at times $t$ and $t+1$ are similar. If they belong to separate train/test sets, overfitting will occur. The accuracy in the simple random ratio-based splitting test set is almost the same as in the train set.
%
%

Moreover, such notations as "a ratio-based split based on subject" are insufficient because the accuracy varies greatly depending on how the test set is selected. For example, in the UCI HAR dataset, subject 14 is an anomaly, and the accuracy changes depending on whether subject 14 is included in the test set. If the train/test set split is unclear, an overfit result can be generated by including the anomaly in the train set. Train and test sets, and also a validation set if possible, must be defined clearly, such as by subject IDs, trial IDs, and file names.

Few studies in the field of IMU-based HAR field satisfy these perspectives and allow direct comparisons. One rare exception where multiple studies can be directly compared is a group of studies using the benchmark setup of the UCI HAR dataset. This dataset is provided in a pre-segmented form, and the train/test set split is subject-based and clearly defined in the subject ID.


Another valid comparison environment is the iSPLInception Benchmark setup. iSPLInception Benchmark setup defines the segmentation settings, and train/test set splits as subject-based, trial-based, or file-based for UCI HAR, PAMAP2\cite{pamap2}, OPPORTUNITY\cite{opportunity}, and Daphnet\cite{daphnet}, considering class sample ratios. Most importantly, the source code of the data handling is opened.


Therefore, as direct comparisons, we use studies that are basically evaluated in the UCI HAR. Studies evaluated in the iSPLInception Benchmark setup will also be compared. However, as described later in Subsection \ref{sec:ispl_bm_issue}, the iSPLInception Benchmark setup has problems with time warping and dirty segmentation, both of which have a small impact on PAMAP2 and Daphnet but a massive impact on OPPORTUNITY. At least for OPPORTUNITY, the iSPLInception Benchmark setup should not be used. Therefore, in this paper, we define a new benchmark setup, called the IMU-based HAR Benchmark, and make it open, to solve these problems and to help future studies.

\subsubsection{LOSO CV}

As for the train/test set split, leave-one-subject-out cross-validation (LOSO CV) is also valid since it clearly determines the train/test set split. However, unfortunately, segmentation is often unclear/different in many studies, and so direct comparison is difficult.


Another problem with LOSO CV is that it increases both the parameter search and evaluation times. Currently, most datasets in the IMU-based HAR domain consist of about ten subjects, an amount that rises up to 50 for a big dataset, in the IMU-based HAR domain. However, this is very small compared to image datasets. If possible, the datasets of the IMU-based HAR domain should be comprised of 1,000 or 10,000 subjects to consider the individual differences. This idea is one of the future works of the IMU-based HAR domain. However, when the dataset size increases, the LOSO CV will cause an evaluation time problem. If increases of evaluation times are acceptable, it is better to evaluate the generality of methods using different datasets that have various types of tasks. In other words, such evaluation should be done on multiple datasets in a well-designed benchmark setup. Therefore, in this paper, we did not select the LOSO CV for our evaluation setup.

\subsubsection{Re-verifiability}
Unfortunately, unlike the image recognition domain, many IMU-based HARs are closed sources and cannot be re-verified. Only a few can be verified, such as DeepConvLSTM and iSPLInception. This situation is unhealthy. For healthier evaluations, the results should be eliminated that others cannot replicate. Such re-verifiability is especially important when identification accuracy approaches 100\% and when tiny improvements are being compared.

\subsubsection{Metrics}
Many studies use accuracy, F1 scores, and weighted F1 scores (wf1) as representative values for the performance of other classifiers. However, F1 scores do not match well for multiclass recognition. Wf1's value is almost the same value as its accuracy; it is not a very effective value. Neither should accuracy nor wf1 should not be used, especially for extremely class-imbalanced datasets. For example, the iSPLInception Benchmark setup for the Daphnet dataset has 2103 samples for class 1 and 126 for class 2. Simply answering every sample as class 1 yields an accuracy and wf1 of 92.84\% and 0.9284. Such high values are misleading. On the other hand, the macro F1 score (mf1) shows 0.5 in the same situation. If a representative value were chosen, mf1 is preferable. If the imbalance is not too extreme, then the accuracy comparison is acceptable. However, in most cases, mf1 shows similar values of accuracy.\footnote{They can be recalculated if the confusion matrix is provided in terms of numbers rather than percentages. Confusion matrices presented as percentages are only valid on one axis. In addition, since this approach basically hides information, it is detrimental and must not be used unless the class is balanced.}

In the datasets discussed so far, OPPORTUNITY, which excludes the null labels, UCI HAR, and PAMAP2 are class-imbalanced but not excessively. Therefore, accuracy, which is used in many studies, can be used for direct comparison. On the other hand, mf1 should be used for Daphnet.



\subsubsection{Summary}
Based on the above discussion, this study mainly compares and discusses studies using the UCI HAR benchmark setup. We also compare the results of several algorithms evaluated with iSPLInception-like benchmark setups called IMU-based HAR Benchmark setups. 

\subsection{IMU-based HAR evaluated by UCI HAR}

As shown above, Table \ref{tab:UCIHAR-comp} summarizes the accuracy of the studies that use the UCI HAR benchmark setup, including the rTsfNet results.

Until around 2015, IMU-based HARs were dominated by handcrafted features, such as statistical features, and traditional ML, such as HMM and SVM. For example, C.A Ronao et al. \cite{doi:10.1177/1550147716683687} achieved 93.18\% accuracy in the UCI HAR benchmark setup with a Hierarchical Continuous HMM. D. Anguita et al. \cite{ucihar} showed 96.37\% accuracy as the reference performance of the UCI HAR dataset with SVM and 561 handcrafted features. 


With the subsequent significant development of DNNs in the image recognition domain, IMU-based HAR has also become DNN-based. C.A. Ronao et al. \cite{RONAO2016235} achieved 94.79\% accuracy in the UCI HAR benchmark setup with CNN. Some studies have been combining CNN and RNN. K. Xia et al. \cite{9043535} achieved 95.14\% accuracy with a combination of LSTM and CNN. L. Lu et al. \cite{9802089} achieved 96.67\% accuracy by combining LSTM and GRU with a multichannel stream approach. Some other studies use Attention, which is a key mechanism of the Transformer's great success with LLMs.
Z.N. Khan et al.\cite{KHAN2021107671} achieved 95.40\% accuracy with a combination of CNN and Attention. 


Some studies have used DNNs and handcrafted features together. For example, C.A. Ronao et al. \cite{RONAO2016235} achieved 95.75\% accuracy by inputting frequency features derived by FFT into a CNN. A. Ignatov \cite{IGNATOV2018915} achieved 96.06\% accuracy by concatenating handcrafted features after feature extraction in a CNN. They also got 97.63\% accuracy by performing data centering, but this method is not treated in this paper because it uses information from its test dataset.

The combination of DL, ML, and handcrafted features shows the highest performance among related studies in Table \ref{tab:UCIHAR-comp}. W. Jiang et al. \cite{10.1145/2733373.2806333} achieved 97.59\% accuracy with a combination of FFT, CNN, and SVM. 


As described above, the importance of the methods using handcrafted features is clear, since their accuracy surpasses the methods using state-of-the-art DNN mechanisms.

\subsection{Re-verifiable Works}\label{sec:ispl_bm_issue}
Unfortunately, the related studies described in the previous subsection are unverifiable, unlike the image recognition domain. DeepConvLSTM and iSPLInception are rare studies whose sources are publicly available and can be re-verified.

DeepConvLSTM is a combined CNN and LSTM model that has been evaluated on the OPPORTUNITY dataset. 
Benchmark setups, which are also available in the public source code, are clearly defined for both segmentation and train/test set splits. 
However, using DeepConvLSTM for direct comparison poses several issues that must be confronted.

The first concerns the handling of NaN values. 
%
In the benchmark setup of DeepConvLSTM, this NaN value is complemented linearly without any time limit. Therefore, a variation rule can be specifically created in the sensor values, an advantageous result for interpretation by RNNs and other methods.\footnote{The OPPORTUNITY dataset has a continuous and long-lasting high occurrence of NaN values specific to certain behaviors and sensors in the 12 accelerometers.} 

The second issue is the identification target difference issue. DeepConvLSTM's identification target is the last label of the segment, a strategy that is different from such common tasks as estimating modes. Therefore, we do not use this setup or perform a direct comparison with DeepConvLSTM.\footnote{In addition, when concatenating multiple trials, the segmentation does not consider the breaks, and time warps occur within some segmentations. However, this issue is a tiny amount, and so the impact is minimal.}


The iSPLInception, which is a model that applies the inception \cite{Szegedy_2015_CVPR} structure proposed in image recognition to HAR, is evaluated using the following four datasets: UCI HAR, a dataset of basic behaviors under the single sensor and simple measurement conditions; PAMAP2, a dataset of basic behaviors under multi-sensor and complex conditions; OPPORTUNITY, a dataset of such daily behaviors as opening a door acquired with multi-sensors, and Daphnet, a dataset that targets the freezing phenomenon of Parkinson's disease, whose capture is challenging to capture with accelerometers compared to basic behavior. The published source code includes the benchmark setups, and both segmentation and train/test set splits and the validation set split, are clearly defined.

However, since the iSPLInception Benchmark setup defines a different split method for the UCI HAR than the original train/test set split, it cannot be used for a direct comparison with other studies validated using UCI HAR. In addition, this benchmark setup has time warps in the segmentation due to the simple exclusion of samples containing NaN values and null labels, and no consideration of the breaks among subjects, trials, and files in the datasets except for the UCI HAR. 

Segmentation affected by simply excluding samples containing NaN values and null labels in PAMAP2 and Daphnet is also minimal and has little impact. Segmentation involving inter-user and inter-trial breaks is also negligible and has little impact on all the datasets. 

Compared to these datasets, the OPPORTUNITY dataset significantly impacts the issue described above. The OPPORTUNITY dataset almost always has samples labeled null between each activity. In addition, as shown in Table \ref{tab:opp_len_summay}, a majority of its activities end in fewer than 90 samples, although the iSPLInception Benchmark setup divides OPPORTUNITY into 90 samples. Therefore, 76.63\% of the segmentations have at least one time warp. Such dirty segmentation is detrimental to discriminators that consider order, signal frequency, etc. 

To solve the issues described above, this study defined an IMU-based HAR Benchmark setup based on the iSPLInception Benchmark setup with an updated NaN value handling and segmentation method, as shown in Subsection \ref{sec:imu_based_har_benchmark}. Although the IMU-based HAR Benchmark and the iSPLInception Benchmark setups are not strictly comparable, the PAMAP2 and Daphnet data sets are less affected by the modifications, and so we use the values before and after the modifications are used for comparison and verification in this paper. On the other hand, since OPPORTUNITY is affected by a large number of modifications, we do not directly compare it. rTsfNet's performance in the iSPLInception Benchmark setup for OPPORTUNITY is presented for comparison, although its advisory values should not be compared in future studies. 

\begin{table}
\caption{Activity length summary of OPPORTUNITY}
\label{tab:opp_len_summay}
\centering
\begin{tabular}{l r}
\toprule
length & \% less than N \\
\midrule
30 & 1.72 \\
32 & 2.75 \\
45 & 10.65 \\
60 & 26.46 \\
75 & 54.98 \\
90 & 76.63 \\
\bottomrule
\end{tabular}
\end{table}

\section{rTsfNet}\label{sec:rtsfnet}
\begin{figure}[tb]
  \centering
  \includegraphics[width=\linewidth]{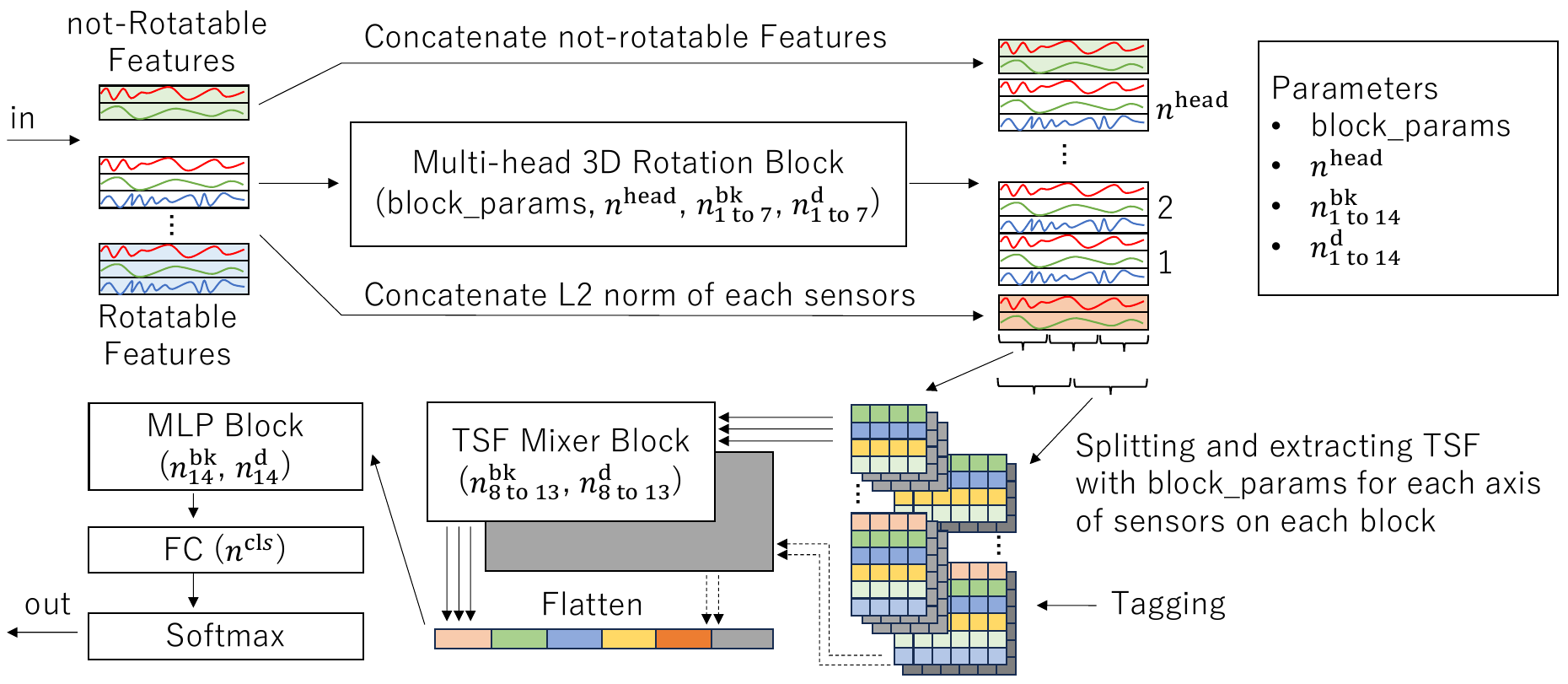}
  \caption{rTsfNet}
  \label{fig:rTsfNet}
  \Description{Overview of the network structure of rTsfNet described in Subsection \ref{sec:rTsfNet_main}}
\end{figure}

\begin{figure}[tb]
  \centering
  \includegraphics[width=0.8\linewidth]{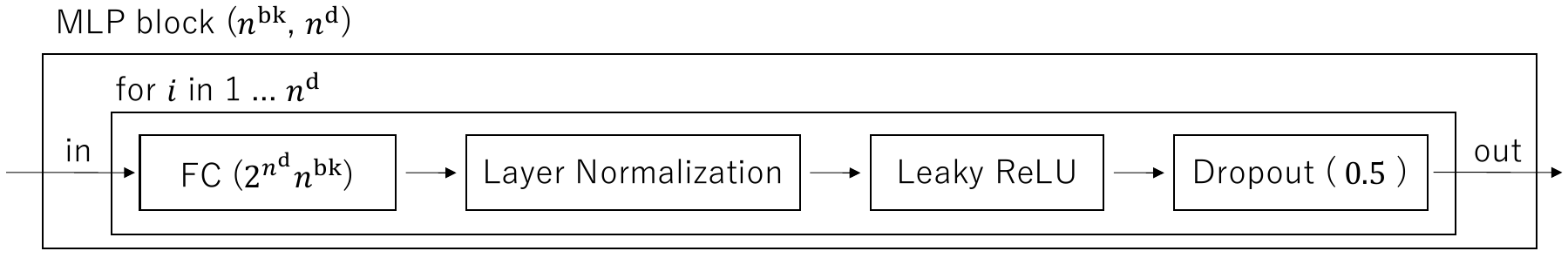}
  \caption{MLP Block}
  \label{fig:mlp}
  \Description{Overview of the MLP Block of rTsfNet described in Subsection \ref{sec:rTsfNet_mlp}}
\end{figure}

\begin{figure}[tb]
  \centering
  \includegraphics[width=0.7\linewidth]{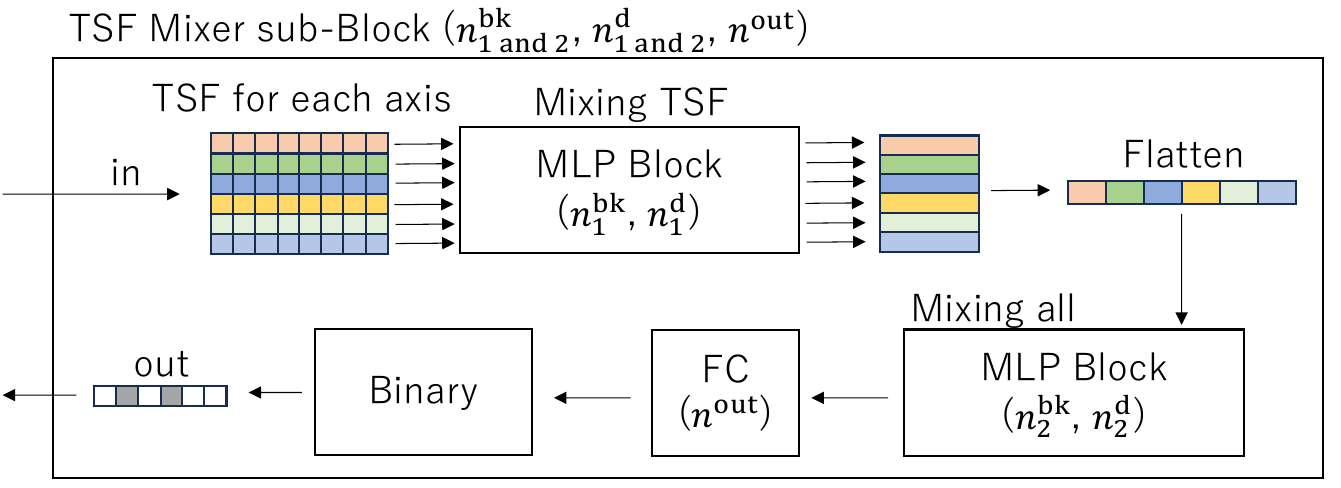}
  \caption{Tsf Mixer sub-Block}
  \label{fig:tsf_m_s}
  \Description{Overview of the TSF Mixer sub-Block of rTsfNet described in Subsection \ref{sec:rTsfNet_tsf_mixer_sub}}
\end{figure}

\begin{figure}[tb]
  \centering
  \includegraphics[width=0.8\linewidth]{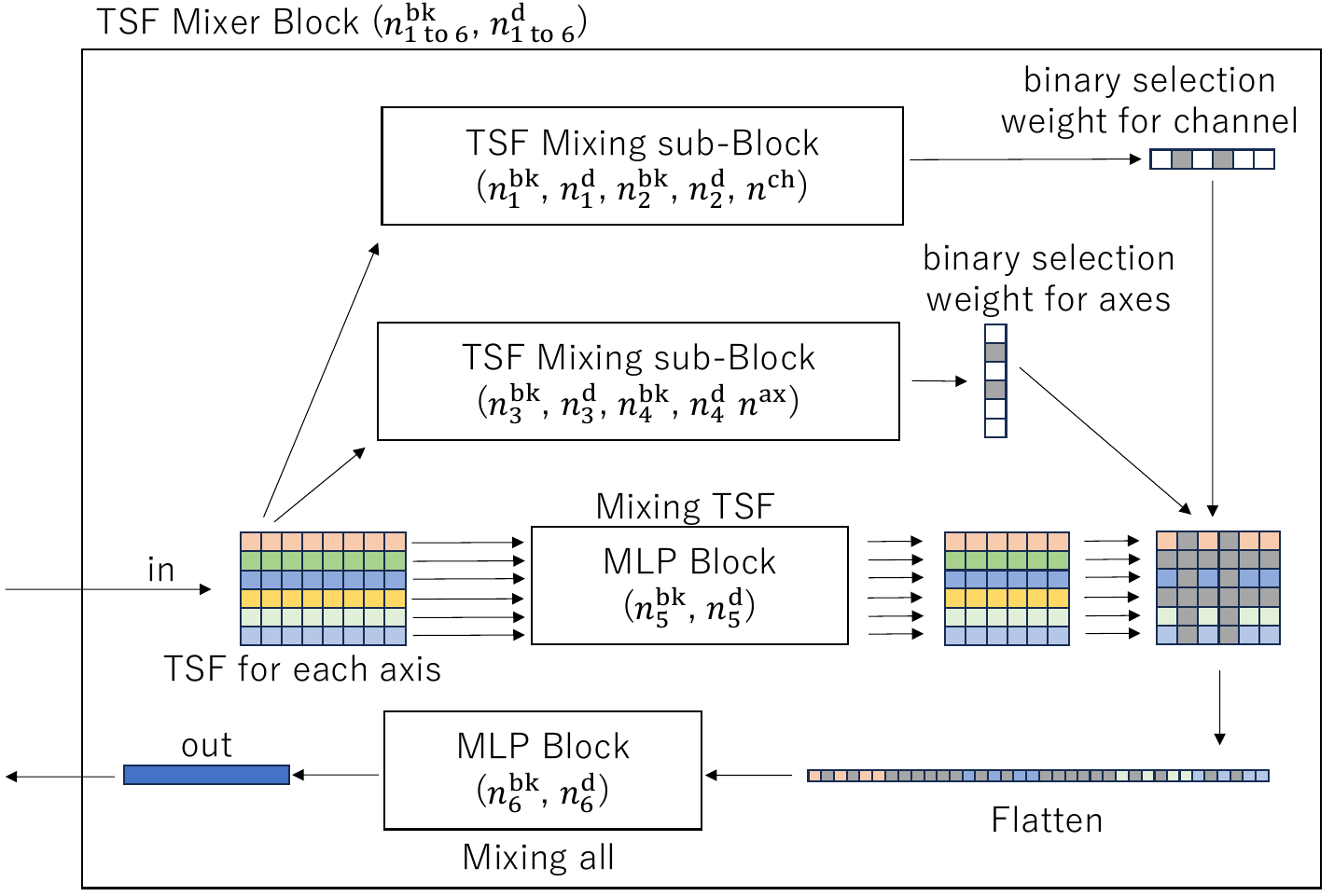}
  \caption{Tsf Mixer Block}
  \label{fig:tsf_m}
  \Description{Overview of the TSF Mixer Block of rTsfNet described in Subsection \ref{sec:rTsfNet_tsf_mixer}}
\end{figure}

\begin{figure}[tb]
  \centering
  \includegraphics[width=\linewidth]{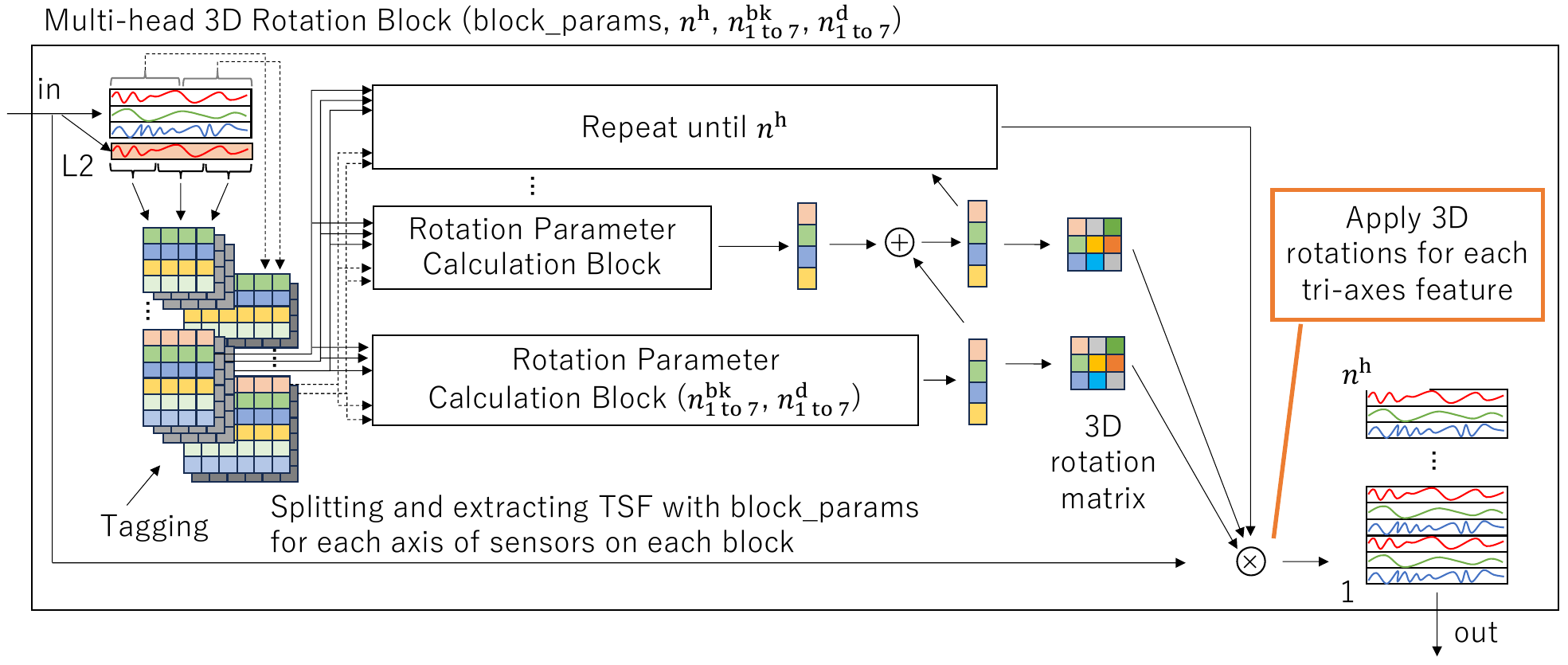}
  \caption{Multi-head 3D Rotation Block}
  \label{fig:mh3dr}
  \Description{Overview of the Multi-head 3D Rotation Block of rTsfNet described in Subsection \ref{sec:rTsfNet_mh3dr}}
\end{figure}

\begin{figure}[tb]
  \centering
  \includegraphics[width=\linewidth]{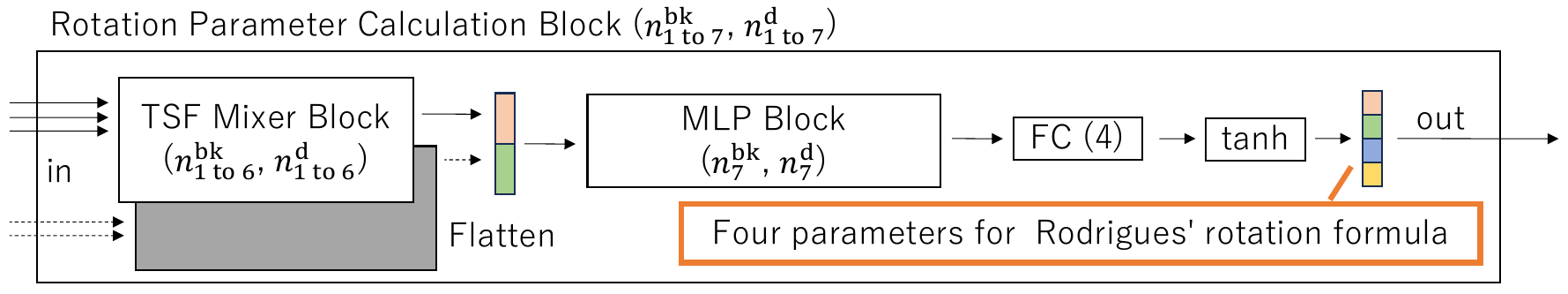}
  \caption{Rotation Parameter Calculation Block}
  \label{fig:rpc_blk}
  \Description{Overview of the Rotation Parameter Calculation Block of rTsfNet described in Subsection \ref{sec:rTsfNet_mh3dr}}
\end{figure}

rTsfNet selects multiple 3D bases in the DNN and derives and extracts time series features from the data in them. An overview is illustrated in Figure \ref{fig:rTsfNet}. The network structure proposed in this paper does not use Residual, SE, Attention, LSTM, ensemble, and so on. We dare to propose it as a basic structure for other networks like CNN. Figures \ref{fig:rTsfNet} to \ref{fig:rpc_blk} show the overall picture.

In rTsfNet, the values from 3-axis sensors that can rotate in 3D are treated separately from other sensor values. The values of 3-axis sensors are subjected to multiple rotations in the Multi-head 3D Rotation Block. Then, we concatenate the L2 norm of the values of the 3-axis sensors that can rotate 3D, the sensor values after the rotation, and the values of the sensors that cannot rotate 3D and extract a time series feature from each axis. Mixed features are extracted from the time series features by the TSF Mixer Block, and the MLP Block performs the final identification.

For clarity, the following description is given in reverse order, starting with the most minor parts.

\subsection{MLP Block}\label{sec:rTsfNet_mlp}
The structure of an MLP Block, which is the basic structural element used in various parts of rTsfNet, is shown in Figure \ref{fig:mlp}. The number of stages of full-connection (FC) layers is $n$, and the number of kernels in the final FC layer is $n^{\mathrm{{\textbf b}ase {\textbf k}ernel}}$. Each FC layer has $n^{\mathrm{bk}} \times 2^{n-i}$ kernels where $i$ shows its distance from the final layer. The output of the FC layers is activated with LeakyReLU after applying Layer Normalization. Next, a 50\% dropout is applied.

\subsection{TSF Mixer sub-Block}\label{sec:rTsfNet_tsf_mixer_sub}


The TSF Mixer sub-Block (Figure \ref{fig:tsf_m_s}) is used in the TSF Mixer Block described in the next section. It has the structure of a TSF Mixer Block without the axis-wise and the channel-wise binary selections. It receives as input the data in which the TSFs for each axis are stored. First, an MLP Block is applied to extract the features within each axis. In this MLP Block, the weight is shared by every axis. The features derived from each axis by the MLP Block are serialized, and another MLP Block derives the final features.

\subsection{TSF Mixer Block}\label{sec:rTsfNet_tsf_mixer}

The TSF Mixer Block (Figure \ref{fig:tsf_m}) is a TSF Mixer sub-Block with axis-wise and channel-wise binary selections.
The binary selection weights are calculated by the TSF Mixer sub-Blocks with input data.
The calculated weights are applied to the mainstream after the axis-wise features are computed. Then we selectively set the values of the specific axes and channels to 0.
This binary selection adaptively eliminates unnecessary axes from subsequent calculations among the axes augmented by the Multi-head 3D Rotation Block described below. Similarly, unnecessary TSF combinations are adaptively eliminated from subsequent calculations.

\subsection{Multi-head 3D Rotation Block}\label{sec:rTsfNet_mh3dr}
The Multi-head 3D Rotation Block (Figure \ref{fig:mh3dr}) calculates multiple 3D rotation parameters from the input 3D rotatable 3-axis sensor values and concatenates them after the rotations. First, we calculate and concatenate the L2 norm from the input. Then, we divided it into multiple block sets along the time series and extracted the TSFs from each. The extracted TSFs may be different for each block set. For example, an input with a data length of 128 is divided into a block set that consists of four blocks with a data length of 32 while deriving the mean and variance, and another block set that consists of one block with a data length of 128 while deriving the frequency feature. After the TSF extraction, tags are added to each axis. The tags consist of sensor location ID, sensor type, and axis type; they are simple integers, like 1, 2, and 3. All the derived TSFs are input into a single Rotation Parameter Calculation Block (Figure \ref{fig:rpc_blk}) to obtain Rodrigues' rotation formula parameters.

The Rotation Parameter Calculation Block first extracts the features for each input block set. Each segment in the block set is input to a TSF Mixer Block whose weight is shared within the block set. All the features extracted per block are serialized, and then the MLP Block derives the overall features. The four parameters required by Rodrigues' rotation formula, the XYZ elements of the rotation axis vectors, and a rotation angle are calculated by the FC layer with four kernels. The tanh was the most effective activation function of those parameters. The use of tanh is an empirical conclusion. Probably, the range limitation of tanh makes an effect.

Then the 3D rotation matrix is calculated from the derived Rodrigues' rotation formula parameters. The XYZ elements of the rotation axis vector are normalized to have an L2 norm of 1. Then the values of the 3-axis sensor, which is the input to Multi-head 3D Rotation, are rotated using the calculated 3D rotation matrix. This process is repeated for the number of heads, $n^\mathrm{{\textbf h}ead}$. However, the second and subsequent  Rodrigues' rotation formula parameters are obtained as the sum of those used before. This structure stabilizes the accuracy. The use of the sum is an empirical conclusion.

\subsection{The main part of rTsfNet}\label{sec:rTsfNet_main}


Finally, the main part of rTsfNet, which uses these parts described above, is shown in Figure \ref{fig:rTsfNet}.
In rTsfNet, the values of the 3-axis sensors that can rotate in 3D are treated separately from other values. Those former sensor values are subjected to multiple rotations in the Multi-head 3D Rotation Block. The same calculations as in the initial part of the Multi-head 3D Rotation Block are then performed to concatenate the L2 norm of the values of the 3-axis sensors that can rotate in 3D, the sensor values after the rotation, and the values of the sensors that cannot rotate in 3D. We then divide the data into several block sets along the time series, derive TSFs for each datum, and add a \emph{tag} to each axis. Features are then extracted for each segment set. Each block in the block set is input to a TSF Mixer Block whose weight is shared within the block set. All the features extracted per block are serialized and the overall features are derived in the MLP Block. The identification results are obtained by FC layers with kernels of the number of classes to be identified and the Softmax activation function.


To summarize the structure of rTsfNet, rTsfNet has the following 29 number hyperparameters:
$n^\mathrm{h}$, $n_\mathrm{1 to 14}^\mathrm{bk}$, and $n_\mathrm{1 to 14}^\mathrm{d}$. It also needs block parameters such as block size, overlap size, and what TSFs will be extracted.

\section{Evaluation}\label{sec:eval}

Due to the verifiability issue mentioned in Section \ref{sec:rw}, we mainly performed comparisons using the UCI HAR Benchmark setup. The evaluation results of the iSPLInception-like benchmark setup, called the IMU-based HAR Benchmark setup, for PAMAP2 and Daphnet are also discussed. However, note that no strict comparison is possible due to the changes in the handling of the NaN values and the boundaries between trials. In addition, the evaluation results of the iSPLInception Benchmark setup for the UCI HAR and OPPORTUNITY datasets are also shown as reference values. The identification results for the newly defined IMU-based HAR Benchmark setup with the OPPORTUNITY dataset are also presented for future studies.

\subsection{Meta settings of rTsfNet numeric hyperparameters}\label{sec:param_limitation}
Due to exploration time issues, the following limitations were applied.
\begin{itemize}
    \item $n_1^\mathrm{h} = 4$
    \item $n_i^\mathrm{bk} = n_j^\mathrm{bk}$, where $i = (2,6)$, $j = (9,13)$
    \item $n_i^\mathrm{d} = n_j^\mathrm{d}$, where $i = (2,6)$, $j = (9,13)$
    \item $n_1^\mathrm{bk} = n_3^\mathrm{bk} = n_5^\mathrm{bk} = n_8^\mathrm{bk} = n_{10}^\mathrm{bk} = n_{12}^\mathrm{bk}$
    \item $n_1^\mathrm{d} = n_3^\mathrm{d} = n_5^\mathrm{d} = n_8^\mathrm{d} = n_{10}^\mathrm{d} = n_{12}^\mathrm{d}$
\end{itemize}

Different values were selected for the parameters $n_{4,11}^\mathrm{bk, d}$ because the axis counts to which the calculated weights will be applied are significantly different. For a similar reason, different values are selected for the parameters $n_{7,14}^\mathrm{bk, d}$ because the layer outputs are significantly different.

\subsection{Block sizes}

We used the following two block sizes with identical TSF extraction settings. One is a short block with a block size of about 0.5 seconds. There is no overlap between the blocks. The specific size depends on the sampling rate of the dataset. The other is a long block, where the entire segmentation is just one block. However, no long block was used in the iSPLInception Benchmark setup for OPPORTUNITY because it worsened the result due to the dirty segmentation issue.

\subsection{Time series feature selection}
This paper uses the time series features shown in Table \ref{tab:selected_features} from the selection using genetic algorithms \cite{NSGA-II} and manual examination. The list of the time series features examined is shown in the Appendix. 
The selection was conducted for each dataset mentioned above, although identical features were selected for the final result.
Features that were insufficiently defined in the table are described below.

\begin{table}[tb]
\centering
\caption{Selected time series features}
\label{tab:selected_features}
\begin{tabular}{l l}
\toprule
Description & Definition \\
\midrule
Minimum &  \\
Maximum &  \\
\vspace{0.5mm}Abs. energy & $\sum_{i=1}^{N}{x_i^2}$ \vspace{0.5mm}\\
\vspace{0.5mm}Abs. sum of changes & $\sum_{i=1}^{N}{|x_i-x_{i-1}|}$ \vspace{0.5mm}\\
\vspace{0.5mm}Mean change & $\frac{1}{N}\sum_{i=1}^{N}{(x_i-x_{i-1})}$ \vspace{0.5mm}\\ 
\vspace{0.5mm}Rooted mean squared & $\sqrt{\frac{1}{N}\sum_{i=1}^{N}{x_i^2}}$ \vspace{0.5mm}\\
\midrule
\multicolumn{2}{l}{Count above start/end values of segments} \\
\multicolumn{2}{l}{Number of crossings with 1st/3rd quartile.} \\
\multicolumn{2}{l}{Mean of FFT amplitude ratio} \\
\multicolumn{2}{l}{Skewness of FFT amplitude} \\
\multicolumn{2}{l}{Mean of ac. with lags of multiple of N (only N=1 was used in this study)} \\
\multicolumn{2}{l}{Kurtosis of ac. with lags of multiple of N (only N=1 was used in this study)} \\
\bottomrule
\end{tabular}
\end{table}

\subsubsection{Mean of FFT amplitude ratio}
This ratio is the mean value of normalized frequency amplitude values derived by the FFT. The normalized range is 0 to 1.
\subsubsection{Mean/kurtosis of autocorrelation values with lags of multiples of N}
These features are the mean/kurtosis values of the autocorrelation values with lags of multiples of N, e.g., N = 2, 4, 6, 8 $\cdots$. The definition of the autocorrelation is $\frac{1}{(n-l)\sigma^2} \sum_{t=1}^{n-l}(X_{t}-\mu )(X_{t+l}-\mu)$. Only N=1 was used in this study. The maximum of the lag is limited to half of the segmentation, e.g., 16 if the segment consists of 32 samples.

\subsection{Training proceedure}
In this paper, we trained rTsfNet using the following settings.
\begin{itemize}
\item start learning late: 0.001 
\item max epochs: 350
\item Reduce learning rate on training loss plateau: 20\% decrease after 10 epochs in training loss plateau
\item Early stop on validation loss plateau: 50 epochs
\item Boot strap protection for early stop: 150 epochs
\end{itemize}

The final model was selected from a model with the best validation loss and the model of the final epoch because the model performance might be enhanced after the validation loss faces a plateau if the training loss was improved \cite{batch_lr_eval}.

\subsection{Evaluation with UCI HAR dataset benchmark setups}
This section compares the performance of rTsfNet with other studies on the UCI HAR dataset benchmark setup.

\subsubsection{UCI HAR dataset}
The UCI HAR dataset \cite{ucihar} consists of two trials with 30 subjects each whose ages ranged from 19 and 48. 
A Samsung Galaxy S II smartphone was fixed on the left side of their waists for the first trial and on anywhere on their waists based on their own preferences for the second trial.
The data from a three-axis accelerometer and a three-axis gyroscope were collected at 50 Hz with six activities: walking, walking upstairs, walking downstairs, sitting, standing, and lying.
The data between each activity were removed from the dataset and then segmented into 128 samples with 50\% overlap. 
The 70\% of the subjects were selected for the train set and the rest for the test set.

The UCI HAR dataset satisfies all the benchmarking requirements.

\subsubsection{Performance comparison}

The rTsfNet's parameters for the UCI HAR are shown in Table \ref{tab:rTsfnet_paramters}, as selected by a genetic algorithm \cite{NSGA-II} and manual examination.
However, not all the spaces have been fully explored due to computing times. A great possibility remains that better parameters exist.

\begin{table}[tb]
    \centering
    \caption{The rTsfNet parameters for each dataset on this study}
    \label{tab:rTsfnet_paramters}
    \begin{tabular}{l l l l l l l}
    \toprule
    Parameter & UCI HAR & UCI HAR & PAMAP2 & Daphnet & OPPORTUNITY & OPPORTUNITY\\
              &  & (no-mh-3D-rot.) & & & (iSPL) & \\
    \midrule
    $n^\mathrm{h}$      & 4 & 4 & 4 & 4 & 4 & 4 \\
    \vspace{1mm}$n_{1,3,5,8,10,12}^\mathrm{bk}$ & 128 & 128 & 128 & 128 & 16  & 64 \\ 
    \vspace{1mm}$n_{1,3,5,8,10,12}^\mathrm{d}$  & 2   & 1   & 1   & 1   & 1   & 1 \\
    \vspace{1mm}$n_{2,9}^\mathrm{bk}$           & 128 & 128 & 32  & 16  & 64  & 64 \\  
    \vspace{1mm}$n_{2,9}^\mathrm{d}$            & 3   & 4   & 3   & 1   & 1   & 3 \\
    \vspace{1mm}$n_{6,13}^\mathrm{bk}$          & 64  & 128 & 32  & 32  & 128 & 128 \\  
    \vspace{1mm}$n_{6,13}^\mathrm{d}$           & 1   & 4   & 2   & 1   & 4   & 2 \\
    \vspace{1mm}$n_{4}^\mathrm{bk}$             & 128 & 16  & 32  & 16  & 16  & 64 \\  
    \vspace{1mm}$n_{4}^\mathrm{d}$              & 4   & 3   & 3   & 3   & 3   & 1 \\
    \vspace{1mm}$n_{7}^\mathrm{bk}$             & 16  & 32  & 128 & 128 & 128 & 64 \\  
    \vspace{1mm}$n_{7}^\mathrm{d}$              & 3   & 2   & 2   & 3   & 4   & 2 \\
    \vspace{1mm}$n_{11}^\mathrm{bk}$            & 16  & 64  & 128 & 64  & 16  & 16 \\  
    \vspace{1mm}$n_{11}^\mathrm{d}$             & 4   & 4   & 4   & 2   & 1   & 3  \\
    \vspace{1mm}$n_{14}^\mathrm{bk}$            & 32  & 128 & 128 & 128 & 128 & 128 \\  
    \vspace{1mm}$n_{14}^\mathrm{d}$             & 1   & 1   & 1   & 1   & 3   & 3 \\
    \midrule
    Block size & 32, 128 & 32, 128 & 64, 256 & 32, 192 & 15 & 16, 32 \\
    \bottomrule
    \end{tabular}
\end{table}

As we have already seen, Table \ref{tab:UCIHAR-comp} summarizes the accuracy of the studies using the UCI HAR Benchmark setup, including the rTsfNet results. Table \ref{tab:rTsfNet_UCI HAR_cm} shows a confusion matrix of the rTsfNet results.

rTsfNet achieved the highest accuracy, mf1, and wf1 results among the related studies. The second best result in Table \ref{tab:UCIHAR-comp}, A. Ignatov \cite{IGNATOV2018915}, used data centering of the statistical features on the dataset, including the test set. Their result falls to the seventh place without the test set's information. The third place in Table \ref{tab:UCIHAR-comp}, W. Jiang et al. \cite{10.1145/2733373.2806333}, is an ensemble and boosting approach that combined of 2D FFT, CNN, and SVM. Thus, the highest performance among the related studies with end-to-end learning of DL is only in the fifth place, L. Lu et al. \cite{9802089}. The rTsfNet performance is 1.09 points higher than that. It is a significant improvement.

\begin{table}[tb]
    \centering
    \caption{Confusion Matrix of rTsfNet for UCI HAR}
    \label{tab:rTsfNet_UCI HAR_cm}
    \begin{tabular}{l r r r r r r}
    \toprule
     & A & B & C & D & E & F\\
     \midrule
    A: Walking & 495 & 1 & 0 & 0 & 0 & 0\\
    B: Walking Upstairs & 4 & 466 & 1 & 0 & 0 & 0\\
    C: Walking Downstairs & 2 & 6 & 412 & 0 & 0 & 0\\
    D: Sitting & 0 & 0 & 0 & 462 & 29 & 0\\
    E: Standing & 0 & 0 & 0 & 23 & 509 & 0\\
    F: Laying & 0 & 0 & 0 & 0 & 0 & 537\\
    \bottomrule
    \end{tabular}
\end{table}

\subsubsection{How well did the Multi-head 3D Rotation functon?}\label{sec:how-the-mh3dr-well}
The fourth place position in Table \ref{tab:UCIHAR-comp} is held by rTsfNet without any Multi-head 3D Rotation. This pattern's parameters are shown in Table \ref{tab:rTsfnet_paramters}. This setup is identical as when TSFs are derived in advance and input to the DNN.

Its acc, mf1, and wf1 are 96.71, 0.9675, and 0.9670, respectively. 
This result shows that the effectiveness of combining the network structure of rTsfNet and the selected TSFs even without the Multi-head 3D Rotation.
Moreover, the 97.76, 0.9779, and 0.9776 results for acc, mf1, and wf1 with Multi-head 3D Rotation show the overall improvements from this result. 
Thus, all of the network structures of rTsfNet, the selected TSFs, and Multi-head 3D Rotation are very effective in the IMU-based HAR.

\subsection{Evaluation with PAMAP2, Daphnet, and Opportunity}
To evaluate the method's generality of the method, rTsfNet was checked with PAMAP2, Daphnet, and OPPORTUNITY, all of which are more complex than UCI HAR. PAMAP2 is a dataset of basic behaviors under multi-sensor and complex conditions; OPPORTUNITY is a dataset of such daily behaviors as opening a door acquired with multi-sensors; Daphnet is a dataset targeting the freezing phenomenon of Parkinson's disease, which is more challenging to capture with accelerometers compared to basic behaviors.

The iSPLInception Benchmark setup clearly defines the segmentation and train/test set split for these datasets; however, its benchmark setup has a time warp issue in the segmentation due to the simple exclusion of samples containing NaN values and null labels and segmentation that does not consider the break between trials, as already described in Subsection \ref{sec:ispl_bm_issue}. Therefore, this study defines a new benchmark setup, called the IMU-based HAR Benchmark, with the following changes from the iSPLInception Benchmark setup.

\begin{itemize}
    \item NaN values: linear interpolation up to 0.2 seconds
    \item Trial boundary: no segmentation over the boundary
    \item NULL label boundary: no segmentation over the boundary
\end{itemize}

In addition, with OPPORTUNITY, the segmentation length is shortened from 90 samples to 32 samples due to the activity length issue, also already described in Subsection \ref{sec:ispl_bm_issue}. 

Segmentation involving inter-user and inter-trial breaks is negligible and has little impact on all the datasets. The segmentation affected by simply excluding samples containing NaN values and null labels in PAMAP2 and Daphnet is also minimal and has little impact.
Compared to these datasets, the OPPORTUNITY dataset significantly impacts this issue. 

Thus, although the IMU-based HAR Benchmark and the iSPLInception Benchmark setups are not strictly comparable, the PAMAP2 and Daphnet data sets are less affected by the modifications, and so we use the values before and after the modifications for comparison and verification in this paper. On the other hand, since OPPORTUNITY is affected by a large number of modifications, we did not directly compare it. rTsfNet's results in the iSPLInception Benchmark setup are presented for comparison, but only as a reference value that should not be compared in future studies. 

\subsection{IMU-based HAR Benchmark}\label{sec:imu_based_har_benchmark}
The IMU-based HAR Benchmark is an open-source platform defined in this study. It is forked from the iSPLInception Benchmark setup.
Most setups are identical as the iSPLInception Benchmark. The following are the changes:
\begin{itemize}
    \item NaN values: linear interpolation up to 0.2 seconds
    \item Trial boundary: no segmentation over the boundary
    \item NULL label boundary: no segmentation over the boundary
    \item Segmentation length of OPPORTUNITY: reduced from 90 to 32
    \item train/test set split of UCI HAR: returned to the original UCI HAR
    \item train/validation split of UCI HAR: newly defined.
\end{itemize}

The summary of each dataset's setup is shown in Table \ref{tab:benchmark_setup}.\footnote{Although they were not used in this paper, we implemented supports for WISDM, RealWorld, the NULL label including segmentation for OPPORTUNITY, LOSO CV for datasets, Optuna based parameter optimization, etc., are implemented. Details are available: \anon[\url{https://xxxx.com}]{\url{https://bit.ly/40b7R1C}}.}

\begin{table}[tb]
\centering
\caption{IMU-based Benchmark setup details (+iSPL for OPPORTUNITY)}
\label{tab:benchmark_setup}
\begin{tabular}{l r r r r}
 \toprule
 & UCI HAR & PAMAP2 & OPPORTUNITY & Daphnet \\
 \midrule
Sampling rate & 50 Hz & 100 Hz & 30 Hz & 64 Hz \\
Segment length & 128 & 256 & 32 (iSPL: 90) & 192 \\
Overlap & 50\% & 50\% & 50\% & 50\% \\
Training set & the others & the others & the others & the others \\
Validation set & 7, 22 & 5 & \small S1-ADL1, S3-ADL3,Drill, S4-ADL4 & \small S02R02, S03R03, S05R01 \\
Test set & \small 2,4,9,10,12,13,18,20,24 & 6 & \small S2-ADL2,Drill, S3-ADL1, S4-ADL5 & \small S02R01, S04R01, S05R02 \\
Split rule & subject-based & subject-based & trial-based & trial-based \\
\bottomrule
\end{tabular}
\end{table}

\subsection{PAMAP2}
The PAMAP2 dataset\cite{pamap2} consists of 11 basic and 7 optional activities with 9 subjects and three IMUs recorded at 100 Hz.
IMU's locations are the hand, the chest, and the ankle. 
The IMUs recorded the readings of accelerometers, gyroscopes, magnetometers as well as the temperature and heart rates.
The iSPLInception Benchmark and IMU-based HAR Benchmark setups use the data of accelerometers, gyroscopes, and magnetometers. 

The following are the basic activities: lying, sitting, standing, walking, running, cycling, nordic walking, ascending stairs, descending stairs, vacuum cleaning, and ironing. The optional activities are watching TV, computer work, car driving, folding laundry, house cleaning, playing soccer, and rope jumping. 
The iSPLInception and IMU-based HAR Benchmark setups target the basic activities.

\subsubsection{Daphnet}
The Daphnet dataset\cite{daphnet} is a dataset to benchmark methods used for recognizing the Freezing of Gait (FOG) of Parkinson's disease (PD), which is a sudden and transient inability to walk. It places wearable acceleration sensors placed on the shank, the thigh, and the lower back.
Almost 50\% of patients with advanced Parkinson's disease face the FOG. It can cause of fails, interference
with daily activities, and a significantly impaired quality of life.

This dataset consists of two targets: freeze and no freeze. 
The data were collected from 10 PD patients at a sampling rate of 64 Hz. 
Due to its disease incidence, this dataset is extremely class-imbalanced.
Thus, this is a very difficult dataset to detect the target.

\subsection{OPPORTUNITY}
The Opportunity activity recognition dataset\cite{opportunity} targets 17 naturalistic activities in daily life, such as ``Open Door'', ``Close Door'', ``Clean Table'', ``Drink from Cup'' and so on. It was collected in a sensor-rich environment with 12 subjects.
Due to frequent continuous NaN issues, the iSPLInception Benchmark and IMU-based HAR Benchmark setups used 7 wearable sensors of many: five XSens of the motion jacket, and two Sun SPOTs on the shoes. This dataset's sampling rate is 30 Hz. Four subjects recorded five trials for the activity set and one long remaining run to collect a large number of activity instances.

\subsection{Performance evaluation}

The rTsfNet parameters for PAMAP2, Daphnet, and OPPORTUNITY are shown in Table \ref{tab:rTsfnet_paramters}. They were selected by a genetic algorithm \cite{NSGA-II} and manual examination. However, not all the spaces have been completely explored due to computing times. Perpahs, better parameters will eventually be identified.


The comparison results are shown in Table \ref{tab:three-db-comp}. The result with the UCI HAR dataset on the iSPLInception Benchmark setup is also listed as a reference value. The rTsfNet showed the highest performance for all the datasets. The confusion matrixes of rTsfNet are shown in Tables \ref{tab:cm-pamap2},  \ref{tab:cm-daphnet}, \ref{tab:cm-opportunity-ispl}, and \ref{tab:cm-opportunity}.

The OPPORTUNITY result with the iSPLInception Benchmark setup is lower than 
OPPORTUNITY with the IMU-based HAR Benchmark setup. These two cannot be compared directly; however, the result suggests that the dirty segmentation of the iSPLInception Benchmark setup caused inaccurate classification.

\begin{table}[tb]
\centering
\caption{Comparision on the IMU-based HAR and iSPLInception Benchmark setups}
\label{tab:three-db-comp}
\begin{tabular}{l l l l l l l l l l}
\toprule
 & \multicolumn{3}{l}{PAMAP2} & \multicolumn{3}{l}{Daphnet} & \multicolumn{3}{l}{UCI HAR (iSPL)} \\
& acc & mf1 & wf1 & acc & mf1 & wf1 & acc & mf1 & wf1 \\   
\midrule
rTsfNet & 95.35 & 0.9353 & 0.9545 & 95.65 & 0.7051 & 0.9466 & 97.49** & 0.9749** & 0.9749** \\  
iSPLInception\cite{ispl} & 89.10 & 0.8786 & 0.8821 & 93.52 & 0.6533 & 0.9212 & 95.09 & 0.9499 & 0.9508 \\
CNN\cite{VANKUPPEVELT2020100548}* & 85.79 & 0.8424 & 0.8399 & 92.97 & 0.5455 & 0.9035 & 91.67 & 0.9160 & 0.9159 \\
CNN-LSTM\cite{9065078}* & 88.37 & 0.8687 & 0.8720 & 92.97 & 0.5258 & 0.8993 & 94.48 & 0.9442 & 0.9441 \\
vLSTM\cite{9065078}* & 85.53 & 0.8368 & 0.8487 & 93.22 & 0.5873 & 0.9106 & 90.80 & 0.9077 & 0.9082 \\
sLSTM\cite{9065078}* & 88.42 & 0.8715 & 0.8866 & 87.65 & 0.5598 & 0.8797 & 91.82 & 0.9180 & 0.9174 \\
BiLSTM\cite{9065250}* & 86.98 & 0.8597 & 0.8646 & 92.41 & 0.4803 & 0.8918 & 93.92 & 0.9383 & 0.9390 \\
\bottomrule
\toprule
 &\multicolumn{3}{l}{OPPORTUNITY (iSPL)} & \multicolumn{3}{l}{OPPORTUNITY } \\
 & acc & mf1 & wf1 & acc & mf1 & wf1 \\   
\midrule
rTsfNet & 91.76 & 0.8815 & 0.9177 & 94.65 & 0.9107 & 0.9462 \\
iSPLInception\cite{ispl} & 88.14 & 0.8369 & 0.8811 &  &  &  \\      
CNN\cite{VANKUPPEVELT2020100548}* & 82.24 & 0.7384 & 0.8005 &  &  &  \\      
CNN-LSTM\cite{9065078}* & 81.41 & 0.7375 & 0.8111 &  &  &  \\      
vLSTM\cite{9065078}* & 76.79 & 0.6949 & 0.7676 &  &  &  \\      
sLSTM\cite{9065078}* & 80.82 & 0.7194 & 0.8002 &  &  &  \\      
BiLSTM\cite{9065250}* & 79.90 & 0.7297 & 0.7995 &  &  &  \\      
\bottomrule
\multicolumn{10}{l}{* No official implementation. It is based on the re-implementation of the iSPLIncepton Benchmark.}\\
\multicolumn{10}{l}{** This model was trained with the original train/test split (meaning less training data).}\\
\end{tabular}
\end{table}

\begin{table}[tb]
\centering
\caption{Confusion matrix of the PAMAP2}
\label{tab:cm-pamap2}
\begin{tabular}{l r r r r r r r r r r r}
 \toprule
 & A & B & C & D & E & F & G & H & I & J & K\\
\midrule
A: Lying             & 173 & 0 & 0 & 0 & 0 & 0 & 0 & 2 & 4 & 2 & 0\\
B: Sitting           & 0 & 46 & 0 & 0 & 0 & 0 & 0 & 1 & 1 & 5 & 0\\
C: Standing          & 0 & 1 & 34 & 0 & 0 & 0 & 0 & 0 & 1 & 0 & 0\\
D: Walking           & 0 & 0 & 0 & 195 & 0 & 0 & 0 & 2 & 2 & 0 & 0\\
E: Running           & 0 & 0 & 0 & 2 & 166 & 0 & 0 & 3 & 6 & 0 & 0\\
F: Cycling           & 0 & 0 & 0 & 0 & 0 & 157 & 0 & 0 & 2 & 0 & 0\\
G: Nordic walking    & 0 & 1 & 0 & 1 & 0 & 0 & 204 & 1 & 0 & 0 & 0\\
H: Ascending stairs  & 0 & 0 & 2 & 0 & 0 & 0 & 0 & 90 & 9 & 0 & 0\\
I: Descending stairs & 0 & 0 & 0 & 0 & 0 & 0 & 1 & 2 & 77 & 1 & 4\\
J: Vacuum cleaning   & 0 & 0 & 0 & 1 & 0 & 0 & 5 & 9 & 1 & 147 & 0\\
K: Ironing           & 0 & 0 & 2 & 0 & 0 & 1 & 0 & 1 & 0 & 0 & 289\\
\bottomrule
\end{tabular}
\end{table}

\begin{table}[tb]
\centering
\caption{Confusion matrix of the Daphnet}
\label{tab:cm-daphnet}
\begin{tabular}{l r r }
 \toprule
 & A & B \\
 \midrule
 A: No freeze & 2095 & 8\\
 B: Freeze & 89 & 37\\
\bottomrule
\end{tabular}
\end{table}

\begin{table}[tb]
\centering
\caption{Confusion matrix of the OPPORTUNITY (iSPL)}
\label{tab:cm-opportunity-ispl}
\begin{tabular}{l r r r r r r r r r r r r r r r r r}
\toprule
 & A & B & C & D & E & F & G & H & I & J & K & L & M & N & O & P & Q\\
\midrule
A: Open Door 1      & 68 & 0 & 4 & 0 & 0 & 0 & 0 & 0 & 0 & 0 & 0 & 0 & 0 & 0 & 0 & 4 & 0 \\
B: Open Door 2      & 0 & 62 & 0 & 4 & 0 & 0 & 0 & 0 & 0 & 0 & 0 & 0 & 0 & 0 & 0 & 0 & 0 \\
C: Close Door 1     & 2 & 0 & 72 & 1 & 0 & 0 & 0 & 0 & 0 & 0 & 0 & 0 & 0 & 0 & 0 & 0 & 0 \\
D: Close Door 2     & 1 & 2 & 0 & 61 & 0 & 0 & 0 & 0 & 0 & 0 & 0 & 0 & 0 & 0 & 0 & 0 & 0 \\
E: Open Fridge      & 0 & 0 & 0 & 0 & 73 & 4 & 0 & 0 & 0 & 1 & 0 & 0 & 0 & 0 & 0 & 0 & 0 \\
F: Close Fridge     & 0 & 0 & 0 & 0 & 5 & 66 & 0 & 0 & 0 & 0 & 0 & 0 & 0 & 0 & 0 & 0 & 0 \\
G: Open Dishwasher  & 0 & 0 & 0 & 0 & 2 & 5 & 48 & 4 & 0 & 0 & 0 & 0 & 0 & 0 & 0 & 0 & 0 \\
H: Close Dishwasher & 0 & 0 & 0 & 0 & 1 & 0 & 3 & 46 & 1 & 0 & 0 & 0 & 0 & 0 & 0 & 0 & 0 \\
I: Open Drawer 1    & 0 & 0 & 0 & 0 & 1 & 0 & 0 & 1 & 43 & 5 & 2 & 0 & 0 & 0 & 0 & 0 & 0 \\
J: Close Drawer 1   & 0 & 0 & 0 & 0 & 0 & 2 & 0 & 0 & 1 & 19 & 2 & 1 & 0 & 0 & 0 & 0 & 3 \\
K: Open Drawer 2    & 0 & 0 & 0 & 0 & 0 & 0 & 0 & 0 & 1 & 1 & 29 & 2 & 3 & 0 & 0 & 0 & 0 \\
L: Close Drawer 2   & 0 & 0 & 0 & 0 & 0 & 0 & 0 & 1 & 0 & 2 & 2 & 21 & 1 & 0 & 0 & 0 & 0 \\
M: Open Drawer 3    & 0 & 0 & 0 & 0 & 0 & 0 & 0 & 0 & 0 & 0 & 2 & 3 & 37 & 1 & 0 & 0 & 0 \\
N: Close Drawer 3   & 0 & 0 & 0 & 0 & 0 & 0 & 0 & 0 & 0 & 0 & 2 & 3 & 5 & 38 & 0 & 0 & 0 \\
O: Clean Table      & 0 & 0 & 0 & 0 & 1 & 0 & 0 & 0 & 0 & 0 & 0 & 0 & 0 & 0 & 97 & 0 & 1 \\
P: Drink from Cup   & 0 & 0 & 0 & 0 & 0 & 0 & 1 & 0 & 0 & 0 & 0 & 0 & 0 & 0 & 1 & 260 & 0 \\
Q: Toggle Switch    & 0 & 2 & 0 & 0 & 0 & 0 & 0 & 1 & 0 & 0 & 0 & 0 & 0 & 0 & 0 & 0 & 51 \\
\bottomrule
\end{tabular}
\end{table}

\begin{table}[tb]
\centering
\caption{Confusion matrix of the OPPORTUNITY}
\label{tab:cm-opportunity}
\begin{tabular}{l r r r r r r r r r r r r r r r r r}
\toprule
 & A & B & C & D & E & F & G & H & I & J & K & L & M & N & O & P & Q \\
\midrule
A: Open Door 1      & 146 & 0 & 15 & 0 & 0 & 0 & 0 & 0 & 0 & 0 & 0 & 0 & 0 & 0 & 0 & 4 & 0\\
B: Open Door 2      & 0 & 143 & 0 & 2 & 0 & 0 & 0 & 0 & 0 & 0 & 0 & 0 & 0 & 0 & 0 & 0 & 0\\
C: Close Door 1     & 2 & 0 & 166 & 0 & 0 & 0 & 0 & 0 & 0 & 0 & 0 & 0 & 0 & 0 & 0 & 0 & 0\\
D: Close Door 2     & 0 & 2 & 0 & 151 & 0 & 0 & 0 & 0 & 0 & 0 & 0 & 0 & 0 & 0 & 0 & 0 & 1\\
E: Open Fridge      & 0 & 0 & 0 & 0 & 156 & 6 & 0 & 0 & 2 & 0 & 0 & 0 & 0 & 0 & 0 & 0 & 0\\
F: Close Fridge     & 0 & 0 & 0 & 0 & 7 & 116 & 0 & 2 & 0 & 0 & 0 & 0 & 0 & 0 & 0 & 0 & 0\\
G: {\small Open Dishwasher}  & 0 & 0 & 0 & 0 & 0 & 0 & 105 & 8 & 13 & 0 & 0 & 0 & 0 & 0 & 0 & 0 & 0\\
H: {\small Close Dishwasher} & 0 & 0 & 0 & 0 & 0 & 0 & 11 & 96 & 0 & 0 & 0 & 0 & 0 & 0 & 0 & 0 & 0\\
I: Open Drawer 1    & 0 & 0 & 0 & 0 & 0 & 0 & 0 & 0 & 76 & 0 & 7 & 0 & 0 & 0 & 0 & 0 & 5\\
J: Close Drawer 1   & 0 & 0 & 0 & 0 & 0 & 0 & 1 & 2 & 5 & 27 & 0 & 5 & 0 & 0 & 0 & 0 & 3\\
K: Open Drawer 2    & 0 & 0 & 0 & 0 & 0 & 0 & 1 & 1 & 5 & 0 & 48 & 0 & 0 & 0 & 0 & 0 & 0\\
L: Close Drawer 2   & 0 & 0 & 0 & 0 & 0 & 0 & 0 & 0 & 0 & 6 & 1 & 29 & 0 & 1 & 0 & 0 & 0\\
M: Open Drawer 3    & 0 & 0 & 0 & 0 & 0 & 0 & 0 & 0 & 0 & 0 & 3 & 0 & 65 & 3 & 0 & 0 & 0\\
N: Close Drawer 3   & 0 & 0 & 0 & 0 & 0 & 0 & 0 & 0 & 0 & 0 & 0 & 2 & 1 & 91 & 0 & 0 & 0\\
O: Clean Table      & 0 & 0 & 0 & 0 & 0 & 0 & 0 & 0 & 0 & 0 & 0 & 0 & 0 & 0 & 235 & 0 & 0\\
P: Drink from Cup   & 1 & 0 & 0 & 0 & 0 & 0 & 2 & 0 & 0 & 0 & 0 & 0 & 0 & 0 & 2 & 656 & 0\\
Q: Toggle Switch    & 0 & 0 & 0 & 0 & 0 & 0 & 0 & 0 & 0 & 0 & 0 & 0 & 0 & 0 & 0 & 0 & 99\\
\bottomrule
\end{tabular}
\end{table}

\subsection{Summary}
rTsfNet showed the highest performance for all the datasets: UCI HAR, PAMAP2, Daphnet, and OPPORTUNITY, although each one has a different sensor setup and varying targets.
This result means that rTsfNet's concept is suitable and has generality for IMU-based HAR. 

As discussed in Subsection \ref{sec:how-the-mh3dr-well}, although the combination of rTsfNet's network structure and the selected TSFs is effective even without Multi-head 3D Rotation, using it shows overall improvements from the results obtained by rTsfNet without it. All the network structures of rTsfNet, the selected TSFs, and the Multi-head 3D Rotation are very effective in IMU-based HAR domain.



\section{Potential}\label{sec:potential}

Next, we described rTsfNet's potential.

\begin{description}
    \item[Cooperation with extensions] rTsfNet can probably be improved in combination with such extension structures as Residual, SE, Attention, LSTM, ensemble, and so on. We propose rTsfNet in this paper as a basic structure for other networks like CNN.  

    \item[As feature extractor] Since rTsfNet can be used as a feature extractor. Therefore, for example, it is likely possible that VAE-like networks can improve their performance with the use of rTsfNet as their encoder. 

    \item[TSF implementation] The TSFs described in this paper are only a few proposed by many researchers. rTsfNet's performance can be improved if more complex, suitable, or lightweight TSFs are implemented into the network.

    \item[TSF and parameter selection] The TSFs and parameters of rTsfNet described in this paper were selected by a genetic algorithm \cite{NSGA-II} and manual examination. However, not every space has been fully explored due to the required computing times. Perhaps even better parameters can be identified. In addition, several parameters in this study have the same values to reduce exploration times, as described in subsection \ref{sec:param_limitation}. If such limitations were removed, their performance would be increased. 
    
\end{description}




\section{Conclusion}\label{sec:conclusion}
As a new DNN model for IMU-based HAR, this paper presented rTsfNet, a DNN model with Multi-head 3D Rotation and Time Series Feature Extraction. rTsfNet automatically selects 3D bases from which features should be derived by extracting 3D rotation parameters within the DNN. Time series features (TSFs) (which embody the wisdom of many researchers) are derived for achieving HAR using MLP. 

Although our model does not use CNN, it achieved higher accuracy than the existing models under multiple datasets, which target different activities, under well-managed benchmark conditions.
rTsfNet's concept is suitable and has generality for IMU-based HAR.

As discussed in Subsection \ref{sec:how-the-mh3dr-well}, although the combination of the network structure of rTsfNet and the selected TSFs is effective even without the use of the multi-head 3D rotation, using it shows overall improvements from the results obtained by rTsfNet without it. It means that all of the network structures of rTsfNet, the selected TSFs, and Multi-head 3D rotation are very effective in IMU-based HAR.

As an additional contribution, this study newly defined an IMU-based HAR Benchmark setup for these datasets and created direct comparability for studies that use the benchmark setup. 

The rTsfNet source code and trained models are available at 
\anon[\url{https://xxxx.com}]{\url{https://bit.ly/40b7R1C}}.
The IMU-based HAR Benchmark system is available at \anon[\url{https://xxxx.com}]{\url{https://bit.ly/45OZ1aT}}.

\begin{acks}
This research was partly supported by the \grantsponsor{GS:JSPS}{Japan Society for the Promotion of Science (JSPS)}{https://www.jsps.go.jp/english/} KAKENHI Grant Number \grantnum{GS:JSPS}{21H03481}, \grantsponsor{GS:JST}{Japan Science and Technology Agency (JST)}{https://www.jst.go.jp/EN/} JST-Mirai Program Grant Number \grantnum{GS:JST}{JP21473170}, and the Cabinet Office and Gifu Prefecture for Human Resource Development, and the Research Project on Production Technology for the Aerospace Industry.
The computation was carried out using the General Projects on the supercomputer "Flow" at the Information Technology Center, Nagoya University.
\end{acks}

\bibliographystyle{ACM-Reference-Format}
\bibliography{rtsfnet}

\appendix

\section{Time series features considered in this paper}

The time series features that are used for the genetic algorithm-based selections are shown in Tables \ref{tab:considered_time_series_features_1} and \ref{tab:considered_time_series_features_2}. We omitted the definitions of well-known features.
\begin{table}[b]
\centering
\caption{Considered time series features (1/2)}
\label{tab:considered_time_series_features_1}
\begin{tabular}{l l l}
\toprule
No. & \multicolumn{2}{l}{description}\hspace{0.7\textwidth} \\
\midrule
1 & mean & \\
2 & minimum &  \\
3 & maximum &  \\
4 & quantiles & the 1st, 2nd (median), and 3rd quartile. \\ 
4 & time based quantiles & the values of 25\%, 50\%, and 75\% point along time series. \\ 
5 & skewness & \\
6 & kurtosis & \\
7 & variance & \\
8 & standard deviation & \\
9 & \vspace{0.5mm}rooted mean squared & $\sqrt{\frac{1}{N}\sum_{i=1}^{N}{x_i^2}}$ \vspace{0.5mm}\\
10 & \vspace{0.5mm}mean change & $\frac{1}{N}\sum_{i=1}^{N}{(x_i-x_{i-1})}$ \vspace{0.5mm}\\ 
\bottomrule
\end{tabular}
\end{table}

\begin{table}[tb]
\centering
\caption{Considered time series features (2/2)}
\label{tab:considered_time_series_features_2}
\begin{tabular}{l l l}
\toprule
No. & \multicolumn{2}{l}{description}\hspace{0.7\textwidth} \\
\midrule
11 & \vspace{0.5mm}sum of change & $\sum_{i=1}^{N}{(x_i-x_{i-1})}$ \vspace{0.5mm}\\ 
12 & \vspace{0.5mm}mean abs. change & $\frac{1}{N}\sum_{i=1}^{N}{|x_i-x_{i-1}|}$ \vspace{0.5mm}\\ 
13 & \vspace{0.5mm}abs. energy & $\sum_{i=1}^{N}{x_i^2}$ \vspace{0.5mm}\\
14 & \vspace{0.5mm}abs. sum of changes & $\sum_{i=1}^{N}{|x_i-x_{i-1}|}$ \vspace{0.5mm}\\
15 & \vspace{0.5mm}abs. max & $\max_{1 \le i \le N} |x_i|$ \vspace{0.5mm}\\
16 & \vspace{0.5mm}CID \cite{cid} & $\sqrt{\sum_{i=1}^{N-1}{(x_i-x_{i-1})^2}}$ \vspace{0.5mm}\\
17 & \multicolumn{2}{l}{count above zero} \\
18 & \multicolumn{2}{l}{count above segment's mean} \\
19 & \multicolumn{2}{l}{count above segment's start value} \\
20 & \multicolumn{2}{l}{count above value of 25\% position along segment's time series} \\
21 & \multicolumn{2}{l}{count above value of 50\% position along segment's time series} \\
22 & \multicolumn{2}{l}{count above value of 75\% position along segment's time series} \\
23 & \multicolumn{2}{l}{count above segment's end value} \\
24 & \multicolumn{2}{l}{number of crossings with zero} \\
25 & \multicolumn{2}{l}{number of crossings with segment's mean} \\
26 & \multicolumn{2}{l}{number of crossings with segment's 1st quantile} \\
27 & \multicolumn{2}{l}{number of crossings with segment's 2nd quantile} \\
28 & \multicolumn{2}{l}{number of crossings with segment's 3rt quantile} \\
29 & \multicolumn{2}{l}{number of crossings with segment's start value} \\
30 & \multicolumn{2}{l}{number of crossings with segment's value of 25\% position along segment's time series} \\
31 & \multicolumn{2}{l}{number of crossings with value of 50\% position along segment's time series} \\
32 & \multicolumn{2}{l}{number of crossings with value of 75\% position along segment's time series} \\
33 & \multicolumn{2}{l}{number of crossings with segment's end value} \\
34 & FFT amplitude & \\
35 & FFT amplitude ratio & \\
36 & \multicolumn{2}{l}{mean of FFT amplitude} \\
37 & \multicolumn{2}{l}{variance of FFT amplitude} \\
38 & \multicolumn{2}{l}{skewness of FFT amplitude} \\
39 & \multicolumn{2}{l}{kurtosis of FFT amplitude} \\
40 & \multicolumn{2}{l}{mean of FFT amplitude ratio} \\
41 & \multicolumn{2}{l}{variance of FFT amplitude ratio} \\
42 & \multicolumn{2}{l}{skewness of FFT amplitude ratio} \\
43 & \multicolumn{2}{l}{kurtosis of FFT amplitude ratio} \\
44 & FFT angle & \\
\midrule
45 & \vspace{0.5mm}autocorrelation & $\frac{1}{(n-l)\sigma^2} \sum_{t=1}^{n-l}(X_{t}-\mu )(X_{t+l}-\mu)$ \vspace{0.5mm} \\
\midrule
46 & \multicolumn{2}{l}{mean of autocorrelation with lags of multiple of N} \\
47 & \multicolumn{2}{l}{variance of autocorrelation with lags of multiple of N} \\
48 & \multicolumn{2}{l}{skewness of autocorrelation with lags of multiple of N} \\
49 & \multicolumn{2}{l}{kurtosis of autocorrelation with lags of multiple of N} \\
\bottomrule
\end{tabular}
\end{table}

\end{document}